\documentstyle[prl,epsfig,twocolumn,aps]{revtex}

\begin{document}

\draft

\title{Nonrealistic Behavior of Mean Field Spin Glasses}

\author{C.M.~Newman}
\address{Courant Institute of Mathematical Sciences,
New York University, New York, NY 10012}
\author{D.L.~Stein}
\address{Departments of Physics and Mathematics, University of Arizona,
Tucson, AZ 85721}

\maketitle

\begin{abstract}

We study chaotic size dependence of the low temperature correlations in the
SK spin glass. We prove that as temperature scales to zero with volume, for
any typical coupling realization, the correlations cycle through {\it
every\/} spin configuration in {\it every\/} fixed observation window. This
{\it cannot\/} happen in short-ranged models as there it would mean that
{\it every\/} spin configuration is an infinite-volume ground state.  Its
occurrence in the SK model means that the commonly used `modified
clustering' notion of states sheds little light on the RSB solution of SK,
and conversely, the RSB solution sheds little light on the thermodynamic
structure of EA models.

\end{abstract}


{\it Introduction.}  The Sherrington-Kirkpatrick~(SK) spin
glass~\cite{SK75} is believed to obey the Parisi replica symmetry breaking
(RSB) solution~\cite{P79}, traditionally interpreted in terms of infinitely
many `pure states' and their overlaps~\cite{P83,MPSTV84a,MPSTV84b,MPV87}.
Because pure states --- and their zero-temperature counterparts, ground
states --- are infinite-volume objects, there has been ambiguity in the
interpretation of these results and in their relevance for realistic models
such as Edwards-Anderson (EA)~\cite{EA75}. In~\cite{NS96} and succeeding
papers, we showed that contradictions arise when results for the SK spin
glass are applied to EA.  In this paper, we give a striking example of the
disconnect between short- and infinite-ranged models which shows the
dangers in transporting notions in {\it either\/} direction.

We will consider the SK Ising spin glass for a system of $N$ spins, in zero
external field.  Its Hamiltonian is:
\begin{equation}
\label{eq:SK}
{\cal H}_{\cal J,N}=-{1\over\sqrt{N}}\sum_{1\le i<j\le N} J_{ij} \sigma_i\sigma_j
\end{equation}
where $\sigma_k=\pm 1$ and the couplings $J_{ij}$ are independent,
identically distributed random variables chosen from a Gaussian
distribution with zero mean and variance one.  The $N^{-1/2}$ rescaling in
(\ref{eq:SK}) ensures a sensible thermodynamic limit for free energy per
spin and related quantities.  For a fixed coupling realization, denoted by
${\cal J}$, Eq.~(\ref{eq:SK}) leads to the finite-$N$ Gibbs distribution,
at inverse temperature $\beta$, on spin configurations $\sigma$:
\begin{equation}
\label{eq:rho}
\rho_N(\sigma)=Z_N^{-1}\exp\left[{\beta\over\sqrt{N}}\sum_{1\le i<j\le N}
J_{ij} \sigma_i\sigma_j\right]\, .
\end{equation}
Here $Z_N$ is the usual $N$-spin partition function, and the dependence of
$\rho$ and $Z$ on $\beta$ and ${\cal J}$ has been suppressed.

Previous rigorous results support certain aspects of the SK
high-temperature solution and the Parisi low-temperature one.  Early
results~\cite{ALR87,FZ87} mostly (though not exclusively) concerned the
high-temperature phase.  Pastur and Shcherbina~\cite{PS91} proved
non-self-averaging of overlaps (which already implies more than a single
pair of states), though for a slightly modified Hamiltonian.  They also
proved self-averaging of the free energy density (see
also~\cite{Com98,Zeg98}).  At around the same time the authors proved that
non-self-averaging implied chaotic size dependence of the overlap
function~\cite{NS92}.  More recent work~\cite{GT02,Tal02} has focused on
proving existence of the thermodynamic limit of the free energy density for
almost every coupling realization.  For more results, see~\cite{Tal02}.

{\it Detection of Multiple States.\/} In short-ranged statistical
mechanical models, including the EA spin glass, pure states (and their
zero-temperature counterparts, ground states) are well-defined objects.
Both finite- and infinite-volume `states' are specified by all
$1,2,\ldots$-spin correlation functions at fixed temperature.  A {\it
thermodynamic\/} state is an infinite-volume state satisfying the DLR
equations~\cite{Ruelle,Lanford,Simon,Georg88}, or, equivalently, is one
which is a limit of finite-volume Gibbs
states~\cite{Ruelle,Lanford,Simon,Georg88,Slawny,Dobrushin,vEvH}.  Pure
states are then defined either in terms of clustering properties or,
equivalently, by a ``non-mixture'' requirement.

Although pure states have been widely used in interpreting the meaning of
the RSB solution of the SK model, there is a difficulty: pure states are
defined for a fixed realization of {\it all\/} of the couplings, but in the
SK model the physical couplings $J_{ij}/\sqrt{N}$ scale to zero as
$N\to\infty$.  To address this problem, pure states have been defined in
the SK model using an analogy to the `clustering' property of pure states
in short-ranged models (although the non-mixture definition has also been
used~\cite{P83}).  A putative pure state $\alpha$ in the SK model has been
defined as one satisfying the modified clustering property~\cite{MPV87,BY86}:
\begin{equation}
\label{eq:modclustering}
\langle\sigma_i\sigma_j\rangle^\alpha - 
\langle\sigma_i\rangle^{\alpha}\langle\sigma_j\rangle^{\alpha}\ \to 0 
\qquad {\rm as }\quad N\to\infty\quad ,
\end{equation}
for any fixed pair~$i,j$. 

It is useful to note that some of these difficulties are absent in the
Curie-Weiss model of the uniform ferromagnet, even though physical
couplings scale to zero there also.  But in the ferromagnet, the couplings
``reinforce'' each other, being nonrandom, so it still makes sense to talk
about positive and negative magnetization states in the $N\to\infty$ limit.
This difference between homogeneous and disordered systems has important
consequences.

We will consider the effects of adding a spin to an $N$-spin system, which
simultaneously requires the introduction of $N$ new couplings (the `cavity
method'~\cite{MPV87}).  Rather than attempting to construct individual pure
states directly, we propose a procedure for detecting the {\it presence\/}
of multiple pure states.  Consider the Curie-Weiss ferromagnet, where
addition of a single spin will not substantially alter a given correlation
function.  In an operational sense, one can in principle keep a record, at
fixed $\beta$, of the values of a finite set of even correlation functions
as $N$ grows.  If these values approach a nonzero limit, as in the
mean-field ferromagnet, one can with some justification describe the
low-temperature phase as consisting of a single pair of `pure states', with
each finite-spin Gibbs state $\rho_N$ being an equal mixture of the two.
If, however, the correlation functions persist in changing their values as
$N$ grows, then one can infer that there must exist multiple pairs of pure
states.  That is, the existence of multiple states manifests itself through
the presence of what we have called {\it chaotic size
dependence\/}~(CSD)~\cite{NS92}.

This perspective makes clear that many states should exist in the SK
model. As $N\to\infty$, any specified correlation function cannot depend on
any finite set of couplings, whose magnitudes all scale to zero.  Given
that the new couplings accompanying each additional spin are chosen {\it
independently\/} of the previous set, it naturally follows that below $T_c$
any specified correlation function should not settle down to a limit.

{\it Chaotic Size Dependence.\/} We can further refine this approach to
{\it quantify\/} the number of states for fixed ${\cal J}$.  We first make
precise the notion of CSD.

{\bf Definition.}  For fixed ${\cal J}$ and a sequence of $N$'s chosen
independently of ${\cal J}$, {\it chaotic size dependence\/} is present if
there is no single limit for some fixed correlation function (computed in
the usual way using (\ref{eq:rho})) as $N\to\infty$.

Spin-flip symmetry ensures that all odd-spin correlation functions vanish.
We will therefore focus on even-spin correlation functions. Our first
theorem proves chaotic size dependence in the SK model below some nonzero
temperature (which presumably corresponds to the transition temperature,
and so we call it $T_c$).  Before stating the theorem, we first prove the
intuitive result that, if any two-point correlation function
$\langle\sigma_i\sigma_j\rangle$, at some fixed $\beta < \infty$, has a
limiting value as $N\to\infty$, its value cannot depend on any finite set
of couplings.

To see this, we note that, for any fixed ${\cal J}$,
\begin{equation}
\label{eq:kl0}
\langle\sigma_i
\sigma_j\rangle_N={\langle\sigma_i\sigma_j\rangle_{N,0}+\langle\sigma_i\sigma_j\sigma_k\sigma_l\rangle_{N,0}\tanh(\beta
J_{kl}/\sqrt{N})\over 1+\langle\sigma_k\sigma_l\rangle_{N,0}\tanh(\beta
J_{kl}/\sqrt{N})}\, .
\end{equation}
where $\langle\cdot\rangle_{N,0}$ denotes a thermal average taken with the
coupling $J_{kl}$ set to zero.  (When $(ij)=(kl)$ this formula reduces to
Eq.~(4.3) in \cite{ALR87}.)  In the limit where $\beta J_{kl}/\sqrt{N}\to
0$, this becomes
\begin{equation}
\label{eq:nodep} 
\langle\sigma_i \sigma_j\rangle_N - \langle\sigma_i\sigma_j\rangle_{N,0} = O(\beta J_{kl}/\sqrt{N})
\end{equation} 
This goes to zero for any finite $\beta$ as $N\to\infty$, proving the
result, which can be extended to zero temperature by taking
$\beta\to\infty$ slower than $\sqrt{N}$.  (This point will be discussed
further below.)

{\bf Theorem 1.}  In the Sherrington-Kirkpatrick Ising spin glass, there
exists a temperature $T_c>0$ below which all two-point spin correlations
asymptotically display chaotic size dependence, with probability one in the
coupling realizations ${\cal J}$.

{\bf Proof.}  We consider nonzero temperature; the zero-temperature 
case is handled by letting $\beta \to \infty$ slower than $\sqrt{N}$,
as mentioned above.  
Suppose that, as
$N\to\infty$, a limit exists for some even-spin correlation function, say
$\langle\sigma_1\sigma_2\rangle$, with strictly positive probability in
${\cal J}$; let $c_{12}(\beta)$ be this limiting value. Because
$c_{12}(\beta)$ does not depend on any finite set of couplings, it must be
the same for almost every ${\cal J}$ by the Kolmogorov zero-one law (see,
for example, \cite{Fel71}).  But by doing a gauge transformation in which
$\sigma_2$ and all the couplings connected to it are inverted, one arrives
at a limit $-c_{12}(\beta)$ for the corresponding ${\cal J}'$.  This
violates the constancy of the limit with respect to changes in ${\cal J}$,
unless $c_{12}(\beta)=0$.  The absence of chaotic size dependence for any
even-spin correlation function therefore requires it to vanish as
$N\to\infty$.

We now show that every two-spin correlation function exhibits chaotic
size dependence below some temperature $T_c$.  By the permutation symmetry
of the coupling distribution, if any $m$-spin function has a limit with
strictly positive probability in ${\cal J}$, then {\it all} $m$-spin
functions have limits for a.e.~${\cal J}$.  At nonzero temperature, there
thus remain two possibilities: either all two-point correlation functions
exhibit CSD, or they all have a zero limit.

But the latter violates a simple bound when $\beta$ is 
sufficiently large.  A straightforward
integration by parts for each Gaussian coupling variable gives
\begin{equation}
\label{eq:endens}
e_N=-{\beta\over
2}(1-1/N) \left(1-\overline{\langle\sigma_1\sigma_2\rangle_N^2}\right) \, ,
\end{equation}
where $\overline{[\cdot]}$ indicates an average over all coupling
realizations, and $e_N$ is the averaged energy per spin in the $N$-spin
system.  Because $e_N$ is bounded from below, there must be a temperature
$T_c$ below which $\overline{\langle\sigma_1\sigma_2\rangle_N^2}$ is
bounded away from zero for any $N$ \cite{ALR87}.  Then
$\langle\sigma_1\sigma_2\rangle_N$ cannot converge to 0 as $N\to\infty$
below $T_c$ for a.e.~${\cal J}$, proving the theorem. $\diamond$

The proof of Theorem 1 should be extendable to higher even-spin correlation
functions, though this isn't necessary for what follows.  We note that if
there is a unique state, such as the paramagnetic state above $T_c$, then
there cannot be chaotic size dependence.  The proof of Theorem~1 then requires all
two-point correlation functions to converge to zero, consistent with
(\ref{eq:modclustering}).

{\it Ground States in the SK Model.}  The presence of CSD for all two-point
correlations below $T_c$ already implies the existence of more than a
single pair of spin-reversed states.  The natural followup is to ask, how
many?

Recall that a ground state is a pure state at zero temperature.  Using the
definition of `pure state' in the SK model given in~\cite{MPV87,BY86},
based on the clustering property~Eq.~(\ref{eq:modclustering}), we note that
any pure state must be regarded as a {\it limit\/} of finite-volume Gibbs
states given by Eq.~(\ref{eq:rho}); otherwise Eq.~(\ref{eq:modclustering})
makes no sense.  Because the remainder of the paper focuses on ground
states, we confine our attention to them here.  In a fixed volume with
specified boundary conditions, such as free, a finite-volume ground state
pair is of course simply the spin-reversed pair of spin configurations of
lowest energy.  So, to be consistent with Eq.~(\ref{eq:modclustering}), a
`ground state' in the SK model should be any convergent sequence of
finite-volume ground states; i.e., where every finite set of spins has a
limiting configuration.  We note that this definition is exactly the same
as the standard definition of a ground state in short-ranged
models~\cite{Ruelle,Lanford,Simon,Georg88,Slawny,Dobrushin,vEvH}.

What happens when chaotic size dependence is present?  In that case, there
must exist {\it two or more\/} subsequence limits (along different
increasing sequences of volumes, all presumably ${\cal J}$-dependent), of
finite-volume ground states.  The number of ground states is then at least
as large as the number of distinct limits (it may be larger because of the
flexibility in choice of boundary conditions).  This construction, used
extensively for short-ranged models like EA, can be applied equally well to
the SK model, and allows comparison between the two.

In the proof of Theorem~1, we studied the zero-temperature limit by taking
both $\beta$ and $N$ simultaneously to infinity, with $\beta$ diverging
more slowly than $\sqrt{N}$.  While we did this to simplify the proof, it
can easily be shown for short-ranged models that this method leads to
limiting ground states without having to first take $\beta\to\infty$ for
finite $N$, and then take $N\to\infty$.  We note that if the energy gap
between the ground state and the low-lying excited states, of the type
considered in~\cite{KM00}, remains of order one (that is, does not approach
zero) as $N\to\infty$, as is widely expected~\cite{MPV87}, then it can be
shown that the two approaches to constructing zero-temperature SK ground
states lead to the same outcome.  We therefore will not concern ourselves
further with this distinction, when using the term ground state.

We now prove that the number of ground states is in fact infinite: 

{\bf Theorem 2.}  The SK spin glass has infinitely many distinct
subsequence limits for the collection of all spin correlations (as~$\beta
\to \infty$, $\beta/ \sqrt{N} \to 0$) for a.e.~${\cal J}$.

{\bf Proof.}  Eq.~(\ref{eq:endens}) shows that as $\beta, N \to
\infty$ (with $\beta/ \sqrt{N} \to 0$),
$\overline{\langle\sigma_1\sigma_2\rangle_N^2} \to 1$.  We can then choose
$N_1 << N_2 << \dots$ (in a ${\cal J}$-independent way) so that every
$\langle\sigma_i\sigma_j\rangle_{N_{\ell}}^2 \to 1$ and hence all limits
have $\langle\sigma_i \sigma_j \rangle~=~\pm 1$.

The essential idea we will use is that, if there is only a finite number
$k$ of ground state pairs, then there must be a positive fraction of
2-point correlations $\langle\sigma_i\sigma_j\rangle$ that are the same in
all $k$ pairs.  By interchanging an average over $(i,j)$'s with one over
${\cal J}$'s, it would then follow that for some fixed $(i_0,j_0)$, the
function $\langle\sigma_{i_0}\sigma_{j_0}\rangle$ would be the same for all
ground states in a positive fraction of ${\cal J}$'s, yielding a
contradiction with Theorem~1.

\begin{figure}
\centerline{\epsfig{file=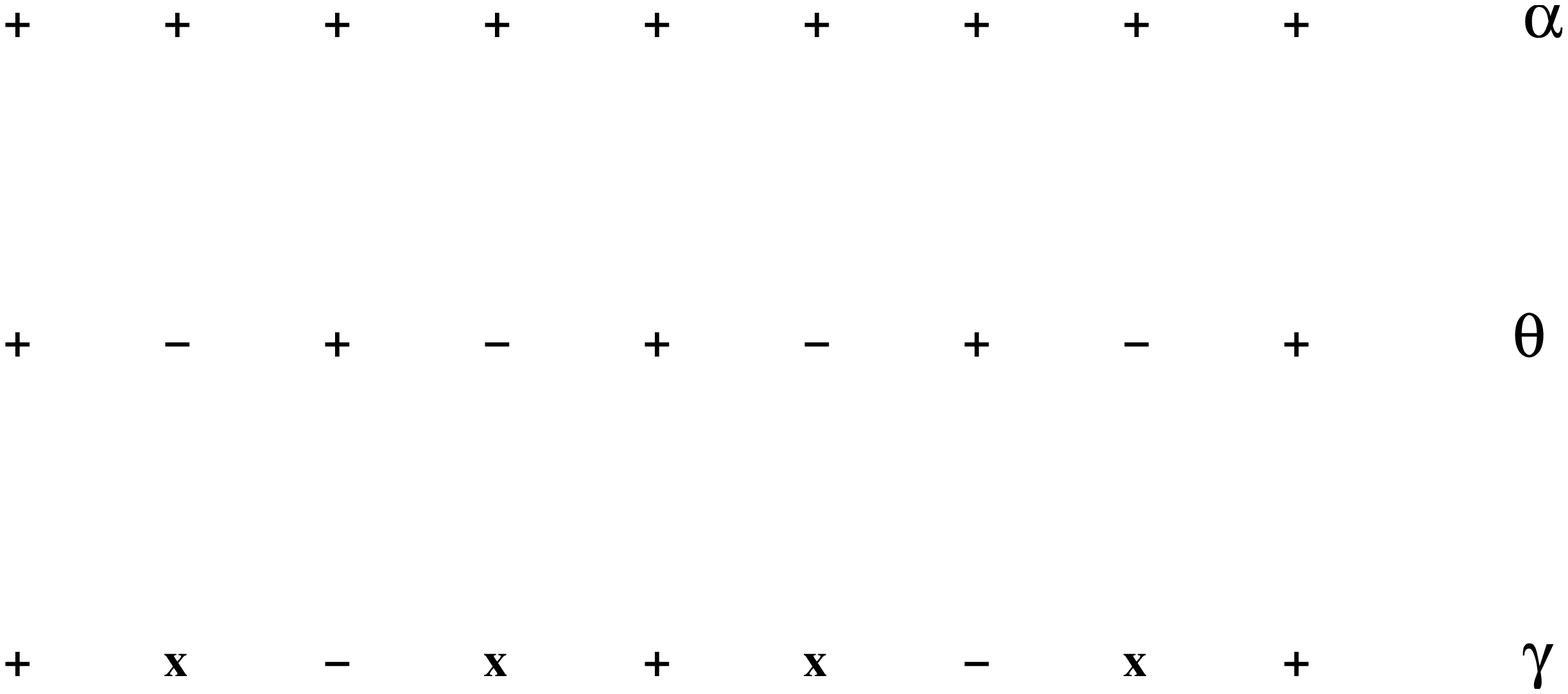,width=2.5in}}
\vspace{+0.1in}
\caption{Spins 1 through 9 in ground states $\alpha$, $\theta$, and $\gamma$
described in the text.  An $x$ indicates that it doesn't matter whether the
spin is $+$ or $-$.}
\vspace{-0.1in}
\label{fig:1}
\end{figure}

By a gauge transformation, we can always choose a ground state --- call it
$\alpha$ --- with all spins $+1$.  Then the maximal disagreement that
ground state $\theta$ can have with $\alpha$ is in half of the 2-point
correlations (cf.~Fig.~1). So $\alpha$ and $\theta$ (and their global
flips) agree on (say) the $\langle\sigma_i\sigma_{i+2}\rangle$
correlations.  If a third ground state $\gamma$ is introduced so that it
maximally disagrees with both $\alpha$ and $\theta$, all three will still
agree on (say) the $\langle\sigma_i\sigma_{i+4}\rangle$ correlations with
$i$ odd.  Continuing, it is easy to see that if there are $k$ ground state
pairs, the $\langle\sigma_i\sigma_{i+2^{k-1}}\rangle,\ldots$ correlations
will all agree for $i$ odd. $\diamond$

{\it Ground State Structure of the SK Model.}  We next ask whether there
exists a countable or uncountable infinity of ground states.  In the
following theorem (which holds for a.e.~fixed ${\cal J}$), we prove not
only that there is an uncountable infinity, but in fact something much
stronger.

{\bf Theorem 3.}  In the Sherrington-Kirkpatrick Ising spin glass, as
$\beta, N \to \infty$ (with $\beta/ \sqrt{N} \to 0$), the signs of $\langle
\sigma_i \sigma_j\rangle$ for $1\leq i,j \leq K$ cycle through the signs of
{\it all\/} $2^K$ spin configurations infinitely often for {\it every\/}
$K$.

{\it Remark.} This is equivalent to the statement that {\it every\/}
infinite-volume spin configuration is a ground state.

{\bf Proof.}  We again focus on volumes $N_1 << N_2 << \dots$ as in the
proof of Theorem~2. For two sizes $M,N$ (with $M << N$), let ${\cal J}^M$
denote all the couplings in the $M$-spin system and ${\cal J}^{N,M}$ denote
all the {\it remaining\/} ones in the $N$-spin system. As $M=M_{\ell}$ and
$N = N_{\ell}$ $\to \infty$, it follows from (\ref{eq:nodep}) that the
limiting ground states are unchanged by setting ${\cal J}^M \equiv 0$
providing $C_M \beta / \sqrt{N} \to 0$ (where $C_M$ is a constant times the
sum of the coupling magnitudes in ${\cal J}^M$). Furthermore, by a simple
gauge transformation argument, when ${\cal J}^M$ is set to zero, the
resulting signs of $\langle \sigma_1 \sigma_j\rangle$ are equally likely
(as ${\cal J}^{N,M}$ varies) to take on the signs of any of the $2^K$ spin
configurations in any $K$-spin system with $K \le M$.

We next choose a sequence of sizes $N'_1 << N'_2 << \dots$ 
from the $N_{\ell}$'s and generate
the full ${\cal J}$ sequentially using ${\cal J}^{N'_1}, {\cal J}^{N'_2,N'_1},
\dots$ so that $C_{N'_{\ell -1}}/\sqrt{N'_{\ell}} \to 0$.
This ensures, with probability one, that for any fixed $K$, the correlation
function values in the $N'_{\ell}$-state, as $\ell \to \infty$, run through
all possible sign configurations infinitely many times. Since this is true
for every choice of $K$, we conclude that, for a typical fixed ${\cal J}$,
the set of all ground states includes {\it every\/} infinite volume spin
configuration.  $\diamond$

{\it Discussion.\/} In this paper we have proved three separate theorems
about the low-temperature phase of the SK model.  The first two together
show that, given the usual notion of ground states, there must be
infinitely many of them.  There is also chaotic size dependence of
finite-$N$ Gibbs states, for all temperatures below $T_c$.  However,
proving infinitely many states is a stronger result than simply showing the
existence of chaotic size dependence, since the latter implies only that
there exists more than a single spin-reversed pair of states.

The idea underlying the proof makes transparent why the SK model displays
many ground states.  It follows from two facts: the independence of the
couplings and their scaling to zero as $N\to\infty$.  The latter condition,
being absent for the EA model, suggests that the presence of many states in
the SK model provides little information on what should be expected in
finite dimensions.

Our third result is that, as $N$ increases, {\it any\/} fixed finite set of
correlation functions cycles through all of its possible spin
configurations infinitely many times.  This striking phenomenon cannot
happen in short-ranged spin glasses in any dimension: it would be
equivalent to {\it every\/} infinite-volume spin configuration being a
ground state for fixed coupling realization.  This is a major qualitative
distinction between SK and EA models.

It goes further.  It demonstrates that the usual, well-defined (for
short-ranged models) notions of pure and ground
states~\cite{Ruelle,Lanford,Simon,Georg88,Slawny,Dobrushin,vEvH}, should
not be carried over to the SK model.  Put another way, our result implies
that any {\it local\/} notion of state, e.g. using correlation functions as
in Eq.~(\ref{eq:modclustering}), leads to absurd conclusions for the SK
model.
 
The standard interpetation~\cite{P83,MPSTV84a,MPSTV84b,MPV87} of the RSB
solution, in terms of infinitely many pure states and their overlaps, even
using definitions modified for the SK model as in
Eq.~(\ref{eq:modclustering}), should therefore perhaps be revisited.  Our
result does not address the idea that there is some more {\it global\/}
notion of states relevant to the SK model, and through which the RSB
solution can be correctly interpreted.  If so, we expect it is unique to
the SK (or related) model(s), since the usual notion of pure and ground
states already suffices for other (short- or long-ranged) systems.

{\it Acknowledgments.\/} This research was partially supported by NSF
Grants DMS-01-02587 (CMN) and DMS-01-02541 (DLS).  DLS thanks N.~Read for a
stimulating correspondence.  We also thank A.~van~Enter for useful comments
on early versions of the manuscript.

\end{document}